\documentclass[twocolumn,journal]{IEEEtran}
\usepackage[LGR,T1]{fontenc}
\usepackage[latin9]{inputenc}
\usepackage{array}
\usepackage{amsbsy}
\usepackage{graphicx}
\usepackage{esint}
\usepackage[unicode=true,
 bookmarks=true,bookmarksnumbered=true,bookmarksopen=true,bookmarksopenlevel=1,
 breaklinks=false,pdfborder={0 0 0},pdfborderstyle={},backref=false,colorlinks=false]
 {hyperref}
\hypersetup{pdftitle={Your Title},
 pdfauthor={Your Name},
 pdfpagelayout=OneColumn, pdfnewwindow=true, pdfstartview=XYZ, plainpages=false}

\usepackage{fancyhdr}
\makeatletter

\DeclareRobustCommand{\greektext}{%
  \fontencoding{LGR}\selectfont\def\encodingdefault{LGR}}
\DeclareRobustCommand{\textgreek}[1]{\leavevmode{\greektext #1}}
\ProvideTextCommand{\~}{LGR}[1]{\char126#1}

\providecommand{\tabularnewline}{\\}

\let\oldforeign@language\foreign@language
\DeclareRobustCommand{\foreign@language}[1]{%
  \lowercase{\oldforeign@language{#1}}}

\usepackage[caption=false,font=footnotesize]{subfig}
\usepackage{cite}

\@ifundefined{showcaptionsetup}{}{%
 \PassOptionsToPackage{caption=false}{subfig}}
\usepackage{subfig}
\makeatother

\begin{document}
\title{Three-Way Serpentine Slow-Wave Structures with Stationary Inflection
Point and Enhanced Interaction Impedance}
\author{Robert~Marosi,~\IEEEmembership{Student Member,~IEEE,}Tarek Mealy~\IEEEmembership{Student Member,~IEEE,}
Alexander~Figotin, and~Filippo~Capolino,~\IEEEmembership{Fellow,~IEEE}\thanks{Robert~Marosi, Tarek\ Mealy, and Filippo\ Capolino are with the
Department of Electrical Engineering and Computer Science, University
of California, Irvine, Irvine, California, e-mail: \protect\href{mailto:rmarosi@uci.edu}{rmarosi@uci.edu}\ \protect\href{mailto:tmealy@uci.edu}{tmealy@uci.edu}~\protect\href{mailto:f.capolino@uci.edu}{f.capolino@uci.edu}.}\thanks{Alexander~Figotin is with the Department of Mathematics, University
of California, Irvine, Irvine, California, e-mail: \protect\href{mailto:afigotin@uci.edu}{afigotin@uci.edu}.}}
\IEEEspecialpapernotice{}
\IEEEaftertitletext{}
\markboth{}{Marosi \MakeLowercase{\emph{et al.}}: Three-Way Serpentine Slow-Wave
Structures with Stationary Inflection Point and Enhanced Interaction
Impedance}
\maketitle

\thispagestyle{fancy}

\begin{abstract}
We introduce two novel variants of the serpentine waveguide slow-wave
structure (SWS), often utilized in millimeter-wave traveling-wave
tubes (TWTs), with an enhanced interaction impedance. Using dispersion
engineering in conjunction with transfer matrix methods, we tune the
guided wavenumber dispersion relation to exhibit stationary inflection
points (SIPs), and also non-stationary, or ``tilted'' inflection
points (TIPs), within the dominant $\mathrm{TE_{10}}$ mode of a rectangular
waveguide. The degeneracy is found below the first upper band-edge
associated with the bandgap where neighboring spatial harmonics meet
in the dispersion of the serpentine waveguide (SWG) which is threaded
by a beam tunnel.

The structure geometries are optimized to be able to achieve an SIP
which allows for three-mode synchronism with an electron beam over
a specified wavenumber interval in the desired Brillouin zone. Full-wave
simulations are used to obtain and verify the existence of the SIP
in the three-way coupled waveguide and fine-tune the geometry such
that a beam would be in synchronism at or near the SIP. The three-way
waveguide SWS exhibits a moderately high Pierce impedance in the vicinity
of a nearly-stationary inflection point, making the SWS geometry potentially
useful for improving the power gain and basic extraction efficiency
of millimeter-wave TWTs. Additionally, the introduced SWS geometries
have directional coupler-like behavior, which enables distributed
power extraction at frequencies near the SIP frequency.
\end{abstract}

\begin{IEEEkeywords}
serpentine waveguide (SWG), serpentine ladder waveguide (SLWG), three-coupled
serpentine waveguide (TCSWG), traveling-wave tube (TWT), millimeter-wave,
dispersion, stationary inflection point (SIP), distributed power extraction
(DPE), interaction impedance.
\end{IEEEkeywords}

\IEEEpeerreviewmaketitle{}

\section{Introduction}

\begin{figure}
\subfloat[]{\includegraphics[width=0.9\columnwidth]{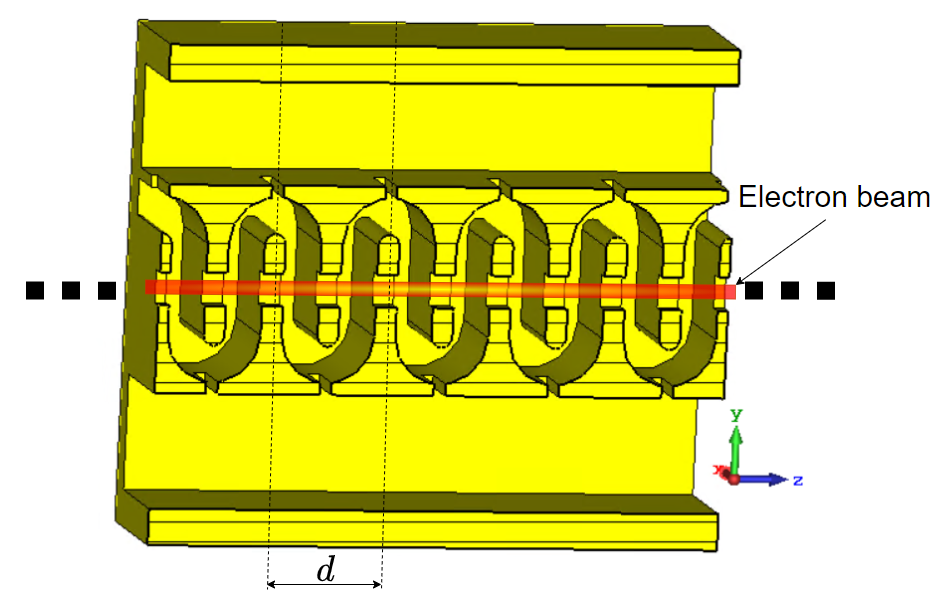}

\label{SLWG_cross_section}}\medskip{}
\subfloat[]{\includegraphics[width=0.9\columnwidth]{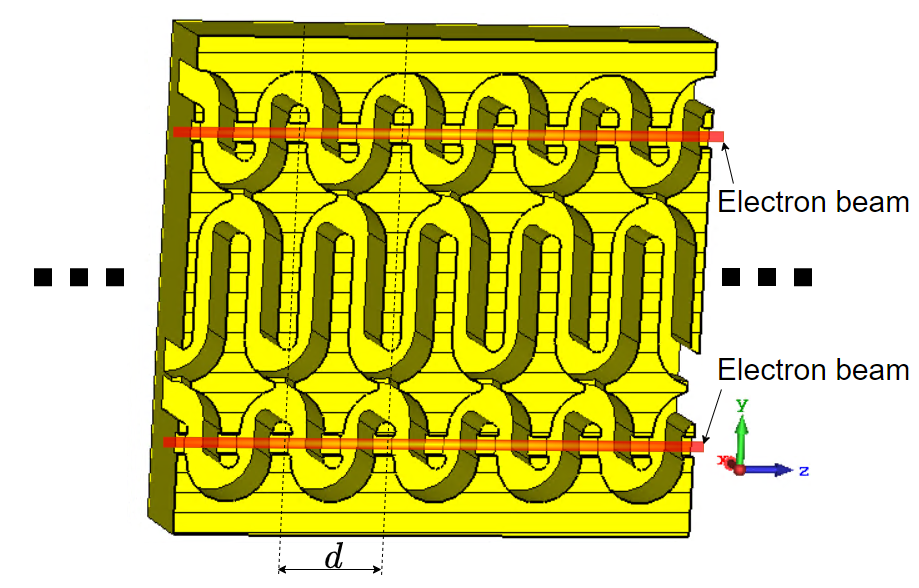}

\label{TCSWG_cross_section}}

\medskip{}
\subfloat[]{\includegraphics[width=0.85\columnwidth]{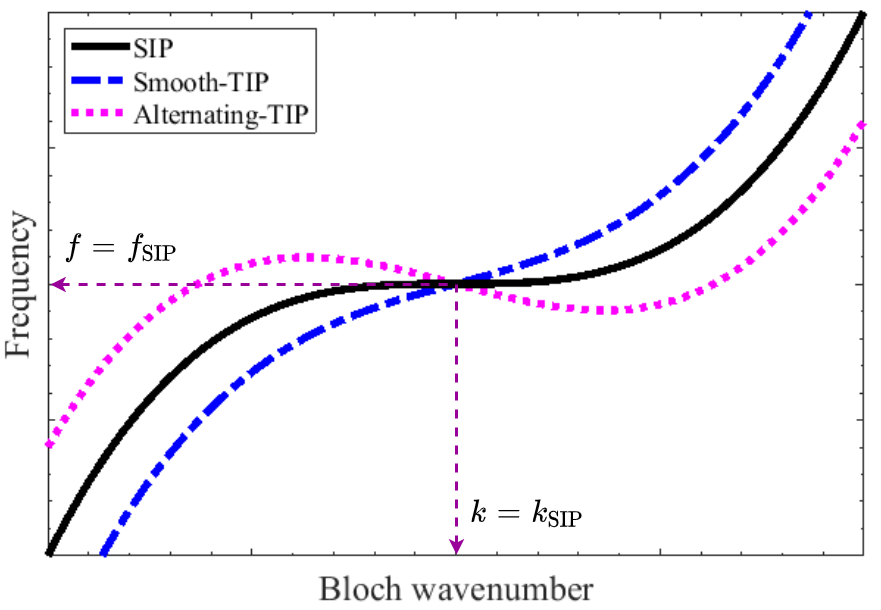}

\label{SIP_illustration}}

\caption{Longitudinal cross-sections in the \emph{y-z} plane showing half of
the periodic metallic structures composed by cascading (a) SLWG unit
cells, and (b) TCSWG unit cells with periodicity $d$. Electron beams
(red) are illustrated in beam tunnels. (c) Example of the dispersion
relation of the mode with purely real wavenumber in the infinite periodic
structure where the SIP occurs (black solid line) at a frequency $f_{\mathrm{SIP}}$
and fundamental wavenumber $k_{\mathrm{SIP}}$ where three modes coalesce
(for clarity, only one mode is shown, the one with purely real wavenumber).
The dispersion relation in the vicinity of an SIP is cubic. The three-way
waveguide dimensions may also be slightly altered to make the SIP
either a smooth-TIP (blue dash-dotted line) or alternating-TIP (magenta
dotted line), rather than an SIP.}
\end{figure}

\IEEEPARstart{T}{raveling-wave} tubes (TWTs) are able to perform
broadband, high-power amplification due to the distributed transfer
of energy from a beam of electrons to guided electromagnetic fields.
There are two primary factors that control the strength of the interaction
between the beam and the electromagnetic field: velocity synchronization
between the beamline and guided modes and beam-wave interaction impedance,
also called Pierce impedance, in the TWT slow wave structure (SWS)
\cite{pierce1951waves}. In S-band helix TWTs, Pierce impedance is
typically on the order of 100 \textgreek{W}, allowing for efficient
basic beam-wave power conversion. However, at millimeter-wave frequencies,
microfabricated slow-wave structures such as the serpentine waveguide
(SWG) typically exhibit Pierce (or interaction) impedance on the order
of 10 \textgreek{W} or smaller, which drastically reduces their basic
conversion efficiency. This is partly due to the fact that serpentine-type
TWTs are typically synchronized to an electron beam in the first or
second Brillouin zone rather than the fundamental Brillouin zone to
avoid the use of relativistic electron beam velocities which require
much larger voltages to accelerate electrons close to the speed of
light \cite{paoloni2021millimeter}. Due to structure geometry that
scales inversely with operating frequency, helix-type TWTs become
extremely difficult to fabricate at millimeter-wave frequencies, making
microfabricated structures like the SWG much more attractive. In this
paper, we propose two new types of dispersion-engineered SWSs based
on the SWG geometry that are capable of exhibiting moderately high
Pierce impedance over narrow bandwidths and can be microfabricated.
These two structure variants, the serpentine ladder waveguide (SLWG)
and the three-coupled serpentine waveguide (TCSWG), with their longitudinal
cross-sections illustrated in Figs. \ref{SLWG_cross_section} and
\ref{TCSWG_cross_section}, respectively, can have their geometries
easily designed to exhibit stationary inflection points (SIPs) or
nearly-stationary tilted inflection points (TIPs), sometimes referred
to as tilted SIPs, at specific frequencies, similarly to the kind
shown in Fig. \ref{SIP_illustration}. Such structures with SIPs are
capable of exhibiting moderately high to very high Pierce (interaction)
impedance comparable to the impedance observed near the band-edge
of SWG SWS, where the group velocity and power flow also vanish. Therefore,
at the frequencies that the electron beam is in synchronism with the
SIP, we expect the proposed SIP SWS to also have a high Pierce gain
parameter, as $C^{3}=Z_{\mathrm{Pierce}}I_{0}/(4V_{0})$, where $I_{0}$
and $V_{0}$ are the average current and equivalent kinetic voltage
of the electron beam, respectively. These structures have potential
use in the design of compact, high efficiency, millimeter-wave TWTs
and backward-wave oscillators (BWOs).

The SIP is a class of modal degeneracy, whereby three eigenmodes coalesce
in both their wavenumbers and eigenvectors (polarization states).
Such modal degeneracies of orders 2, 3, and 4 were originally investigated
by Figotin and Vitebskiy in \cite{figotin2003electromagnetic,figotin2005gigantic,figotin2006frozen,figotin2007slow,figotin2006slow,figotin2011slow}.
The SIP is a particular type of exceptional point of degeneracy (EPD),
and is sometimes called a ``frozen mode'' in literature. Exceptional
points of degeneracy of various orders have been previously investigated
theoretically in gainless and lossless structures operating at both
radio and optical frequencies in \cite{ramezani2014unidirectional,stephanson2008frozen,othman2016giant,almhmadi2019frozen,paul2021frozen,mumcu2009lumped,othman2016theory,burr2013degenerate,othman2016low,nada2017theory,oshmarin2019new,abdelshafy2020distributed,zuboraj2017propagation,yazdi2017new},
and have also been experimentally demonstrated at radio frequency
in \cite{othman2017experimental,chabanov2008strongly,mealy2020general,abdelshafy2018exceptional,oshmarin2021experimental,nada2020frozen}.
In particular, the first experimental demonstration of SIPs at radio
frequencies in a reciprocal three-way waveguide SWS has been performed
in \cite{nada2020frozen}. The slow wave structures we introduce here
are designed to operate with an electron beam synchronized to three
degenerate ``cold'' eigenmodes. That is, the SIP is made to exist
in the ``cold'' dispersion relation, i.e.,before the introduction
of an electron beam, which will perturb the dispersion relation. Note
that this regime of operation is different from the regime of ``exceptional
synchronization'' or ``degenerate synchronization'' studied in
\cite{mealy2021high,mealy2021high2,mealy2019exceptional}, where an
exceptional point of degeneracy is designed to occur in the ``hot''
system, i.e., it is visible only in the hot modal dispersion relation.
These exceptionally synchronized modes of the structure become degenerate
when a synchronized electron beam is coupled to the electromagnetic
modes.

We define a ``waveguide way'' as an individual waveguide component
(rectangular waveguide, SWG, etc.) which is not in cutoff over the
designed operating frequency and can support two electromagnetic modes
(one forward and one backward). A three-way waveguide supports six
modes, when considering propagation in both the $+z$ and $-z$ directions,
i.e., three modes in each direction. The three-way microstrip structure
with SIP in \cite{nada2020frozen} is what inspired the design of
the structures in this paper.

The concept of ``three-mode synchronization regime'' using SIPs
in the cold SWS dispersion of linear-beam tubes was initially proposed
in \cite{yazdi2017new} and was based on ideal transmission lines,
following a multi-transmission line generalization of the Pierce model.
Here, we show for the first time how one can design a realistic serpentine-like
SWS to exhibit a cold SIP at millimeter-wave frequencies.

In Section \ref{sec:Cold-Stationary-Inflection}, we explain the concept
of the SIP, smooth-TIP, and alternating-TIP. In Section \ref{sec:Proposed-Waveguides-exhibiting},
we introduce the two proposed waveguide structures and we describe
our design methodology to introduce SIPs in the three-way coupled
waveguides. In Section \ref{sec:Results}, we show the dispersion,
scattering parameters, and Pierce (interaction) impedance enhancement
for our three-way coupled waveguides. All dimensions used in this
paper are in SI units unless otherwise stated.

\section{Cold Stationary Inflection Points\label{sec:Cold-Stationary-Inflection}}

Stationary inflection points are a special case of eigenmode degeneracy,
whereby three eigenmodes coalesce at a single frequency point in the
dispersion diagram for the modes of a periodic structure, as has been
explored in \cite{figotin2001nonreciprocal,F1,mumcu2005rf,sumetsky2005uniform,stephanson2008frozen,scheuer2011optical,gutman2012slow,apaydin2012experimental,li2017frozen,yazdi2017new,almhmadi2019frozen,gan2019effects}.
On the other hand, a \textit{non-stationary}, or ``tilted'' inflection
point is a single frequency point in the structure's modal dispersion
diagram where three eigenmodes are nearly coalescing, but not perfectly
so. In the structure's modal dispersion diagram, the dispersion relation
local to an SIP or TIP will be cubic in shape. In general, these cubic-shaped
dispersion relations will have an inflection point that occurs at
a frequency-wavenumber combination where the second derivative of
the $\omega-k$ dispersion relation vanishes (i.e. $d^{2}\omega/dk^{2}=0$).
These inflection points are classified here into two kinds: the SIP,
and the non-stationary inflection point, or TIP. The SIP occurs where
both the first and second derivative of the $\omega-k$ dispersion
relation vanish at the same wavenumber, i.e., $d\omega/dk=0$ and
$d^{2}\omega/dk^{2}=0$. For the case of the TIP, only the second
derivative of the $\omega-k$ dispersion relation vanishes at the
inflection point. In other words, the SIP is a special case of TIP.
Because the modal dispersion diagrams for our lossless, reciprocal
structures are symmetric in each Brillouin zone, the usual classification
of non-stationary inflection points as rising or falling is ambiguous.
Every reciprocal waveguide with a TIP in its dispersion relation will
always exhibit two kinds of TIPs, rising and falling, for opposite
signs of $k_{\mathrm{TIP}}$. To remedy this, we futher classify TIPs
into two sub-categories: \textit{smooth}-TIPs which have a non-vanishing
group velocity that does not change sign for wavenumbers slightly
above and below the inflection point, and \textit{alternating}-TIPs
which have a group velocity that alternates in sign as the wavenumber
is swept near the inflection point, as illustrated in Fig. \ref{SIP_illustration}.

We were able to design both the SLWG and TCSWG structures to exhibit
SIPs in their cold dispersion, that is, without an electron beam present.
We stress that this form of modal degeneracy is different from the
hot eigenmode degeneracy, or ``exceptional synchronization'' studied
in \cite{mealy2021high,mealy2021high2}. The coalescence of the cold
eigenmodes at an SIP results in a perfect cubic dispersion relation
local to the SIP in the dispersion diagram,

\begin{equation}
\left(f-f_{\mathrm{SIP}}\right)\simeq h\left(k-k_{\mathrm{SIP}}\right)^{3}.\label{eq:Ideal_SIP_dispersion}
\end{equation}

\noindent When the cold eigenmodes of a structure are nearly-coalescing
at a single wavenumber, and the dispersion relation exhibits a TIP,
the dispersion relation local to the inflection point may be represented
by a depressed cubic function (the quadratic term is suppressed due
to the shift in $k$ by the cubic function's inflection point),

\begin{equation}
\left(f-f_{\mathrm{TIP}}\right)\simeq h\left(k-k_{\mathrm{TIP}}\right)^{3}+s\left(k-k_{\mathrm{TIP}}\right).\label{eq:tilted_SIP_dispersion}
\end{equation}
The parameters $f_{\mathrm{SIP}}\simeq f_{\mathrm{TIP}}$ and $k_{\mathrm{SIP}}\simeq k_{\mathrm{TIP}}$
are the frequency and Floquet-Bloch wavenumber, respectively, at which
the three eigenmodes coalesce or are nearly-coalescing to form an
SIP or a TIP, respectively. We also note that, due to Floquet-Bloch
spatial harmonics, the wavenumbers $k$, $k_{\mathrm{TIP}}$, and
$k_{\mathrm{SIP}}$ in (\ref{eq:Ideal_SIP_dispersion}) and (\ref{eq:tilted_SIP_dispersion})
have a periodicity of $2\pi/d$, where the pitch of the unit cell
is $d$. That is, $k$, $k_{\mathrm{TIP}}$, and $k_{\mathrm{SIP}}$
in these formulae do not necessarily need to be within the fundamental
Brillouin zone. The parameter $h$ is a scalar flatness coefficient
that depends on the strength of eigenmode coupling and $s$ is a scalar
coefficient that affects the ``tilt'' of the TIP, as demonstrated
in Fig. \ref{SIP_illustration}. Tilted inflection points and their
properties, such as improved gain-bandwidth products and power efficiency,
have also been explored in \cite{yazdi2017new}, using a multi-transmission
line generalization of the Pierce model \cite{figotin2013multi,tamma2014extension}.

At an SIP, the dispersion relation local to the inflection point is
a cubic function similar to Eqn. (\ref{eq:tilted_SIP_dispersion})
with $s=0$, as shown in solid black in Fig. \ref{SIP_illustration}.
At, and very close to, an SIP or smooth-TIP, the eigenwaves all propagate
in the same direction. That is, the group velocities of the eigenwaves
do not change sign at frequencies slightly lower and higher than $f_{\mathrm{TIP}}$.
At the inflection point of an SIP, the group velocity becomes zero.
In a smooth-TIP with $s>0$, the group velocity ($d\omega/dk$) at
the inflection point is no longer zero (blue dash-dotted line in Fig.
\ref{SIP_illustration}). Having the electron beam interact with an
TIP with a near-zero positive group velocity (nearly-stationary TIP)
instead of a perfectly-zero group velocity SIP may be preferable for
TWT designs, since the Pierce (interaction) impedance at the frequency
of the inflection point will be sufficiently large but the device
will not become absolutely unstable when the electron beam is introduced.
On the other hand, having an electron beam interact with a TIP that
has negative group velocity is useful for the design of BWOs. We call
this interaction between an electron beam and a cold SIP, a ``three-mode
synchronization'' (see \cite{yazdi2017new} for more details). Beamline
interactions at points of zero group velocity, like the band-edge
\cite{hung2015absolute,zhang2016absolute,antoulinakis2018absolute,ang1998absolute}
or the degenerate band-edge (DBE) \cite{othman2016low,othman2016giant2},
are to be avoided in the design of TWT amplifiers, as they are considered
a source of instability. The alternating-TIP (magenta dotted line
in Fig. \ref{SIP_illustration}) is the second kind of TIP studied
with $s<0$, in which the group velocities of the eigenwaves will
change sign at frequencies slightly lower and higher than $f_{\mathrm{TIP}}$.
If the geometry of a structure can be tuned to exhibit smooth-TIPs
for one set of dimensions and alternating-TIPs for another set of
dimensions, it is expected that such a structure can be made to exhibit
an SIP.

From Pierce theory, the Pierce (interaction) impedance for a specific
Floquet-Bloch spatial harmonic $p$ and specific wavenumber corresponding
to the frequency of interest is defined as

\begin{equation}
Z_{\mathrm{Pierce}}(k_{p})=\frac{\left|E_{z,p}(k)\right|^{2}}{2\mathrm{Re}(k_{p})^{2}P(k)}\label{eq:Zpierce}
\end{equation}
where, $k_{p}=k+2\pi p/d$ is the wavenumber corresponding to the
appropriate \textit{p}th Floquet-Bloch spatial harmonic, $p=0,\pm1,\pm2,...$,
and the wavenumber $k$ is in the fundamental Brillouin zone defined
here as $kd/\pi\in[-1,1]$, i.e., with $p=0$. Furthermore, $\left|E_{z,p}(k)\right|$
is the magnitude of the phasor of the electric field component along
the center of the beam tunnel in the \textit{z} direction for a given
wavenumber and \textit{p}th Floquet-Bloch spatial harmonic, and $P(k)$
is the time-average power flux at the fundamental wavenumber corresponding
to the frequency of interest (the time average power flux is the sum
of power contributions from all spatial harmonics) \cite{gewartowski1965principles_ch10}.
To obtain the magnitude of the axial electric field phasor, corresponding
to the appropriate spatial harmonic, the complex axial electric field
along the beam tunnel axis is decomposed into Floquet-Bloch spatial
harmonics as $E_{z}(z,k)=\sum_{p=-\infty}^{\infty}E_{z,p}(k)e^{-jk_{p}z}$,
where the harmonic weights are computed as $E_{z,p}(k)=\frac{1}{d}\intop_{0}^{d}E_{z}(z,k)e^{jk_{p}z}dz$
\cite{mm2017interaction}. Both the complex axial electric field $E_{z}(z,k)$
and the time-average power flux $P(k)$ through the cross section
of the unit cell are calculated for the cold structure (i.e. without
the electron beam) using the eigenmode solver in CST Studio Suite.
However, one must pay careful attention to how the phase across the
periodic boundaries is defined in the CST model to correctly compute
the interaction impedance. Since the $\exp(j\omega t)$ time convention
is used by CST, the formula for calculating $E_{z,p}(k)$ requires
a delaying phase from the lower periodic boundary to the upper periodic
boundary of the simulated unit cell, i.e. phase of $E_{z}(z,k)$ must
decrease from $z_{min}$ to $z_{max}=z_{min}+d$ for a positive value
of $k$. Conveniently, the electromagnetic energy simulated within
the enclosed vacuum space of the unit cell between periodic boundaries
in the eigenmode solver, which is based on the finite element method
(FEM) implemented in the software CST Studio Suite, is always assumed
to be 1 Joule for each eigenmode solution. Therefore, the power flux
is calculated using the formula $P=(1\,\mathrm{Joule})v_{g}/d$, where
$d$ is the unit cell pitch, and the group velocity $v_{g}=d\omega/dk$
is determined directly from the dispersion diagram via numerical differentiation
(The group velocity is the same at every spatial harmonic).

In order for interaction impedance to be large , the ratio in Eqn.
(\ref{eq:Zpierce}), $\left|E_{z,p}(k)\right|^{2}/P(k)$, must become
large in magnitude or the wavenumber in the denominator must become
very small (i.e., operating closer to the fundamental spatial harmonic).
At a nearly-stationary TIP, which is close to becoming an SIP, the
power flow at the inflection point is indeed smaller than the power
flow of conventional SWG at the same wavenumber of the inflection
point, this because the power flow is proportional to the group velocity.
Assuming that the magnitude of the axial electric field component
is comparable for both cases, one concludes that the Pierce impedance
will be larger for the structure with an inflection point than in
a conventional SWG, at the wavenumber corresponding to the inflection
point.

Due to this phenomena, it is possible to obtain a moderately high,
narrowband Pierce impedance at an SIP or nearly-stationary TIP, which
is several times larger than the Pierce impedance observed in a conventional
serpentine waveguide, as demonstrated in Sec. \ref{sec:Results}.

Conventional TWT SWS exhibit higher symmetries, such as glide symmetry
in the serpentine-type TWT or screw symmetry in the helix-type TWT
\cite{hessel1973propagation,bagheriasl2019bloch}. Nearly parallel
dispersion curves are formed by breaking glide symmetry. Glide symmetry
can be broken in our structures by introducing minor dimensional differences
between two similar waveguide sections. This allows us to readily
tune the tilt of TIPs in our dispersion relation by simply varying
one or more of our structure\textquoteright s dimensions.

\begin{figure}
\subfloat[]{\includegraphics[width=3.5in]{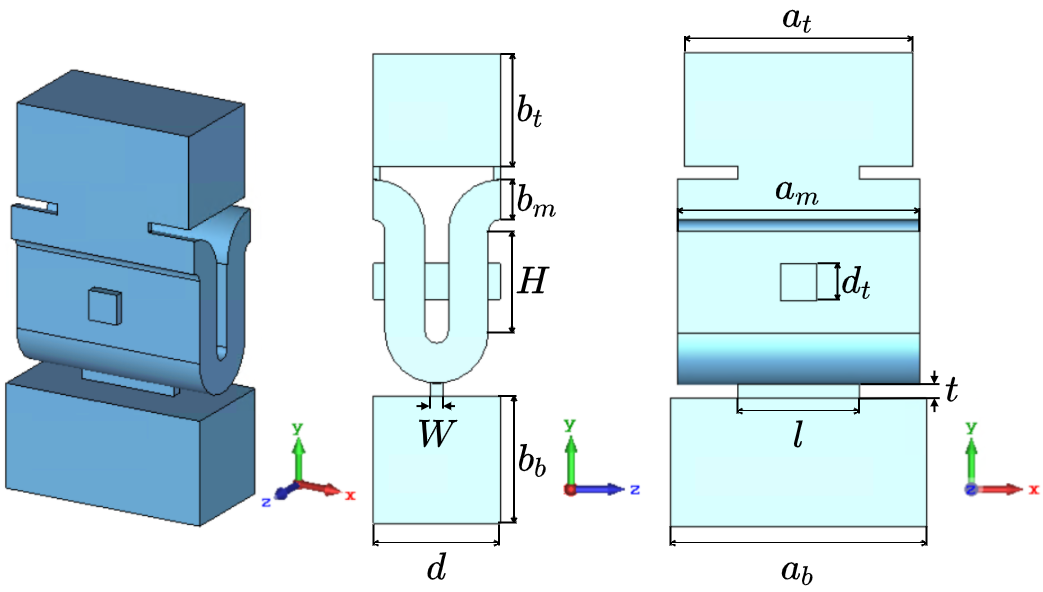}

\label{SLWG_Markup}}

\medskip{}
\subfloat[]{\includegraphics[width=3.5in]{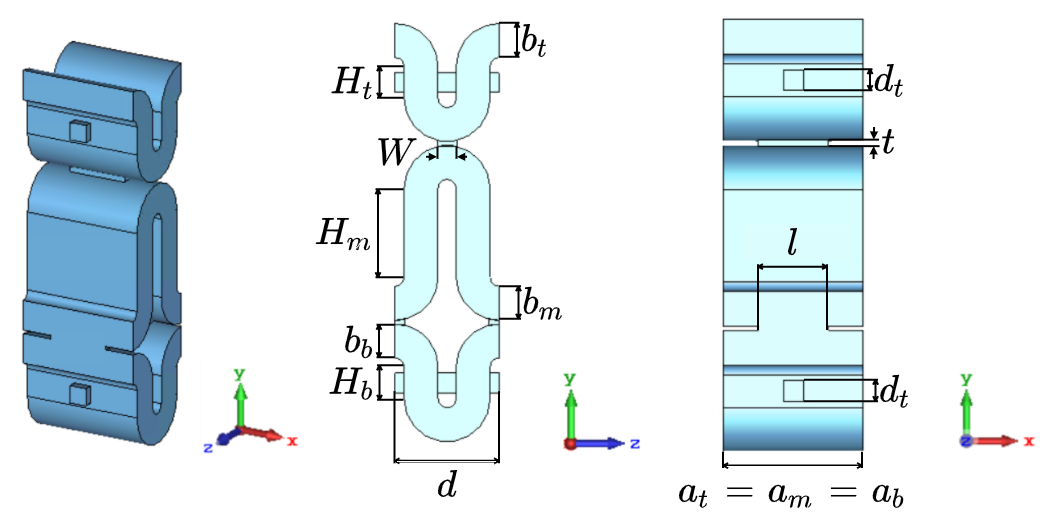}

\label{DualBeamTCSWG_markup}}

\caption{The two unit cell geometries of the two three-way waveguides studied
in this paper. The unit cells are constructed by periodically coupling
three waveguides. Only the vacuum space inside the waveguides is shown
(i.e., no metallic walls are shown). (a) SLWG unit cell with one square
beam tunnel threaded through the middle SWG. The middle SWG is coupled
to two lateral straight waveguides with larger waveguide heights ($b$
dimension) than the middle SWG, (b) Dual-beam TCSWG unit cell with
square beam tunnels threaded through top and bottom SWGs. Background
material is PEC. Waveguide ways have subscripts \emph{t},\emph{ m},
and \emph{b} (top, middle, and bottom) based on their relative positions
along the \textit{y}-axis. Rectangular coupling slots which are in
cutoff are added between middle SWG bends and adjacent waveguide ways.\label{fig:unit_cell_markups}}
\end{figure}

Next, we show the design methodology for two kinds of SWS geometries,
whose unit cells are shown in in Fig. \ref{fig:unit_cell_markups},
that can be dispersion engineered to exhibit SIPs or nearly-stationary
TIPs.

\section{Proposed Waveguides exhibiting SIP\label{sec:Proposed-Waveguides-exhibiting}}

\subsection{Serpentine Ladder Waveguide (SLWG)}

The SLWG was our first attempt at obtaining a SWG-like structure which
is capable of exhibiting an SIP and has its geometry shown in Fig.
\ref{SLWG_Markup}. As we will show in our dispersion diagrams in
Section \ref{sec:Results}, the SLWG structure can be potentially
designed to operate as a BWO due to the backward waves that are exhibited
where the beamline interacts with an SIP or smooth-TIP. This behavior
is also illustrated by the intersection of the inflection point (solid
brown curve) and the beamline (dashed red line) in the dispersion
diagram of Fig. \ref{SLWG_dispersion_approx}. Furthermore, since
the guided electromagnetic modes of the SLWG structure are distributed
over a larger cross section than a conventional SWG due to the two
lateral waveguides that couple to the middle SWG, the power handling
capability of this structure may be enhanced. The SLWG structure is
a serpentine waveguide SWS which is sandwiched between two straight,
parallel rectangular waveguides with similar broad wall dimensions.
A single beam tunnel is through the the center of the SWG structure.
The structure resembles a ladder due to the rung-like appearance of
transverse serpentine sections running between the parallel straight
waveguides. The parallel waveguides are each weakly coupled periodically
to the SWG at the middle of each bend by small rectangular slots,
as shown in the longitudinal cross section of Fig. \ref{SLWG_cross_section}
and in the dimensional markup of the unit cell, Fig. \ref{SLWG_Markup}.
This waveguide structure was inspired by a similar microstrip three-way
periodic structure which also exhibits SIPs \cite{nada2020frozen}.
However, our structures are only able to exhibit SIPs in the fundamental
$\mathrm{TE_{10}}$rectangular waveguide mode by breaking glide symmetry
at this time.

The dispersion curves corresponding to the $\mathrm{TE_{10}}$ mode
of the individual parallel rectangular waveguides must be bent and
vertically shifted in order to obtain an SIP or TIP of a desired tilt
and frequency in the SLWG dispersion relation, as will be further
described in Sec. \ref{subsec:Design-methodology}.

Introducing small dimensional differences between the top and bottom
parallel waveguide ways (i.e., changing the broad wall dimensions
$a_{t}$ and $a_{b}$ as shown in Fig. \ref{SLWG_Markup}) or introducing
metallic obstacles of different dimensions in the top and bottom parallel
waveguides breaks glide symmetry. Breaking glide-symmetry provides
a simple route to achieve a cold SIP in the structure\textquoteright s
dispersion relation once periodic coupling is introduced between individual-waveguide
modes. These differences between the top and bottom waveguides directly
control the tilt of the TIP. In general, breaking glide symmetry is
not a necessary condition to have an SIP, as three-way waveguide structures
with unbroken glide-symmetry have been previously shown to exhibit
SIPs \cite{nada2020frozen}. However, here it is has been found to
be convenient to achieve and manipulate SIPs in structures with broken
glide-symmetry, as the SIP can occur within the fundamental $\mathrm{TE_{10}}$
mode of the SWG below upper band-edge associated to the bandgap of
the SWG which occurs at $k=2\pi p/d$ .

\subsection{Three Coupled Serpentine Waveguide (TCSWG)}

The three-coupled serpentine waveguide TCSWG structure is our second
example of a serpentine-type SWS that is capable of exhibiting SIPs
in its dispersion relation, and has its geometry shown in Fig. \ref{DualBeamTCSWG_markup}.
The TCSWG seems to be better suited for use in a TWT than the SLWG
structure, since the example we show tends to exhibit forward waves
in proximity to the synchronization point where the beamline interacts
with an SIP or a smooth-TIP, as we will show in our dispersion diagrams
in Section \ref{sec:Results}. The TCSWG structure is constructed
similarly to the SLWG structure. However, the top and bottom rectangular
waveguides which sandwich the center SWG are also made to be serpentine
in shape, giving them a longer path length and similar dispersion
shape to that of the center SWG, as shown in the longitudinal cross-section
of Fig. \ref{TCSWG_cross_section} and in the dimensional markup of
the unit cell in Fig. \ref{DualBeamTCSWG_markup}. No periodic obstacles
or broad wall dimension variations are required, as the mean path
lengths of each individual waveguide way may be altered to break glide-symmetry
and obtain an SIP. In our TCSWG structure, the top, bottom, and middle
SWGs each have the same pitch and broad wall dimension. The straight
section height of each SWG way ($H_{t}$, $H_{m}$, $H_{b}$) is varied
to alter the shape of its respective dispersion curve, allowing the
prementioned SIP conditions to occur. The tilt of the TIP is controlled
by the size of the coupling slots placed between bends of adjacent
SWSs, as well as minor path length differences (between the serpentine
height dimensions $H_{t}$ and $H_{b}$) introduced to break glide
symmetry. We have also found that if the path length of the middle
SWG, mainly controlled by $H_{m}$, is made to be a scalar multiple
of the top or bottom path length that is greater than two, it is possible
to have multiple SIPs or TIPs in synchronism with the beamline at
different frequencies, though we do not show it in this paper.

Instead of inserting a beam tunnel only in the center SWG, it is also
possible to add beam tunnels to the center of the top and bottom SWGs.
This makes the structure operate with two beams propagating parallel
to each other, provided that the two beams are not too close together
and an external magnetic field can be used to confine both beams.
This dual-beam structure can potentially benefit from increased power
output at the SIP/TIP due to beamline synchronism in both the top
and bottom SWG sections.

\subsection{Design methodology \label{subsec:Design-methodology}}

A minimum of three ways is required to obtain an SIP in a reciprocal,
lossless, cold structure. This is because the SIP is a synchronous
coalescence of three eigenmodes. In order to design three-way serpentine
waveguide SWSs that are both consistently synchronized to a beamline
and exhibit SIPs, we utilize a design methodology based on the work
of \cite{nguyen2014design}. We use a dispersion approximation for
the initial design of both the individual straight rectangular waveguide
and serpentine waveguide ways. The design process begins by selecting
a fixed center operating frequency, spatial harmonic number $p$,
cell pitch $d$, and average electron beam velocity $u_{0}$ determined
from the cathode-anode voltage of an electron gun. The full-cell pitch
$d$ in our work is chosen to be equal to $\lambda_{g}/4$, where
$\lambda_{g}=2\pi/\left(k_{0}\sqrt{1-c/(2af_{\mathrm{center}})^{2}}\right)$
is the guided wavelength at the center operating frequency within
the SWG containing the beam tunnel, where $k_{0}$ is the free space
wavenumber, $c$ is the velocity of light in free space, $a$ is the
broad-wall dimension of the individual waveguide cross-section as
shown in Fig. \ref{fig:unit_cell_markups}, and $f_{\mathrm{center}}$
is the center operating frequency .

Starting with the average beam velocity, $u_{0}$, the beamline's
linear relation (neglecting space charge effects) between the frequency
and average electronic phase constant , $\beta_{0}$, is
\begin{equation}
\beta_{0}=\frac{2\pi f}{u_{0}}.\label{eq:beamline}
\end{equation}

\noindent Then, the dispersion for the $p^{th}$ spatial harmonics
of the individual serpentine and/or straight waveguide sections is
calculated from the relation found in \cite{nguyen2014design} 
\begin{equation}
f=f_{c}\sqrt{1+\left(\frac{a}{L}\right)^{2}\left(\frac{kd}{\pi}-2p\right)^{2}},\label{eq:parabolic_dispersion_approx}
\end{equation}
where, $f_{c}=c/(2a)$ is the cutoff frequency of the rectangular
waveguide cross-section and $L=2H+\pi d/2$ is the mean path length
of the individual uncoupled waveguide section within the unit cell,
as can be observed in the serpentine waveguide sections of Fig. \ref{fig:unit_cell_markups}.
To model the dispersion of a straight rectangular waveguide sections
of Fig. \ref{SLWG_Markup}, the path length simply becomes equal to
the pitch of the unit cell, $L=d$. We describe our structure using
a full unit cell notation (of period \textit{d}) rather than the half
unit cell notation (of period \textit{d}/2) commonly used in literature.
The half-cell notation is often used because the beam ``sees'' two
beam tunnel intersections per geometric period of the full unit cell,
which only differ by a sense-inversion of $E_{z}$ fields at each
consecutive beam tunnel intersection. The primary difference is that
the path lengths of each individual waveguide of our full unit cell
are twice as long as they would be in the half-cell notation. Much
of the fundamental spatial harmonic of the full-cell dispersion diagram
lies above the light-line of $k_{0}=\pm\omega\sqrt{\mu_{0}\epsilon_{0}}$
and cannot be utilized for amplification without the use of a relativistic
beam velocity \cite{paoloni2021millimeter}. Because the full-cell
notation is being used in this paper rather than the half-cell notation,
the additional $\pi$ phase shift considered in some other papers
such as \cite{nguyen2014design,booske2005accurate,xu2020theory,robertson201971}
is no longer needed, so the term $2p+1$ in \cite{nguyen2014design}
has been replaced with $2p$ in (\ref{eq:parabolic_dispersion_approx}).
As long as the coupling between waveguide sections in each unit cell
is weak, this dispersion relation will serve as a reasonably accurate
approximation of actual dispersion below the first $k=2\pi p/d$ bandgap,
which occurs at the intersection of two neighboring spatial harmonic
curves corresponding to the same individual waveguide way which contains
the beam tunnel.

\begin{figure}
\subfloat[]{\includegraphics[width=0.9\columnwidth]{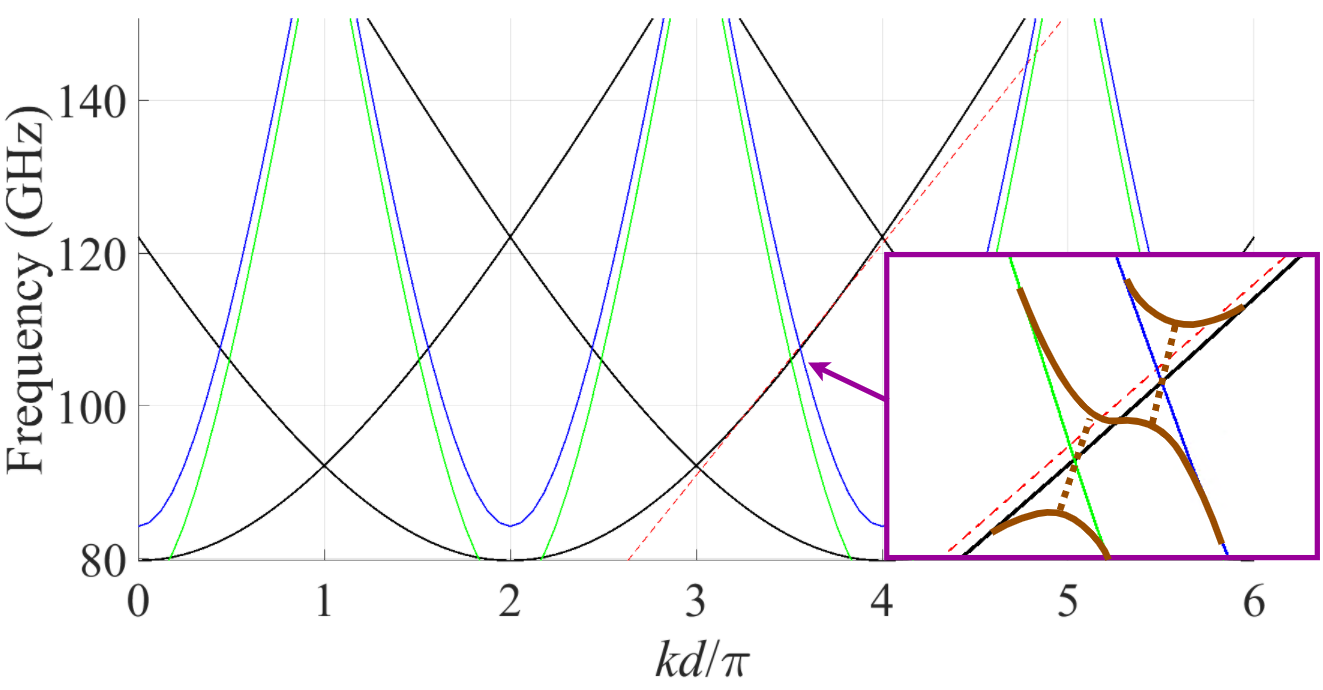}

\label{SLWG_dispersion_approx}}\medskip{}
\subfloat[]{\includegraphics[width=0.9\columnwidth]{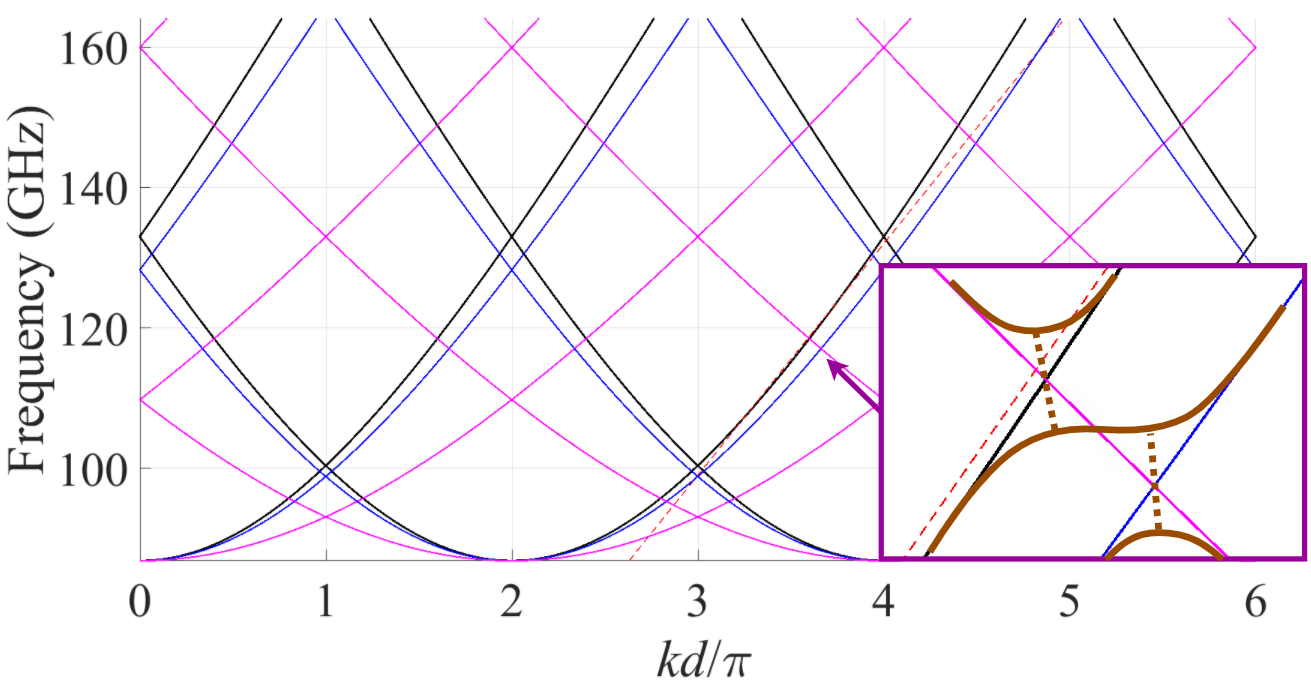}

\label{DualBeamTCSWG_dispersion_approx}}

\caption{Dispersion for the individual (uncoupled) waveguides in (a) SLWG structure
with a small difference between the broad-wall widths of top and bottom
waveguides $\Delta a=\left|a_{t}-a_{b}\right|$ (blue and green curves
for the straight waveguides, black for the SWG, and dashed red curve
for the beamline) and (b) TCSWG structure with a small difference
between serpentine heights of top and bottom SWGs $\Delta H=\left|H_{t}-H_{b}\right|$
(blue and black curves for the top and bottom SWGs, magenta for the
center SWG, and dashed thin red curve for the beamline). Each solid
curve corresponds to the dispersion of an individual-waveguide way
based on Eqn. \ref{eq:parabolic_dispersion_approx}. The width and
cutoff frequency of each dispersion curve can be set by adjusting
the corresponding geometry for the individual-waveguide way. Insets
of figures (a) and (b) illustrate SIP formation when coupling is introduced
(thick brown curves, dotted brown lines for tilted bandgaps). Introducing
coupling between the waveguides causes bandgaps to form at each dispersion
curve intersection. If two nearly-parallel individual-waveguide dispersion
curves cross over a third individual-waveguide dispersion curve, it
is possible to obtain an SIP. The SIP or TIP occurs where the lower
band-edge of one bandgap meets the upper band-edge of an adjacent
bandgap. Tilted inflection point \textquotedblleft tilt\textquotedblright{}
may be be adjusted by varying coupling strength or by adjusting the
distance between parallel individual-waveguide dispersion curves.
Dimensions for this example are provided in Appendix \ref{sec:Dimensions-used-in}.\label{fig:dispersion_approx}}
\end{figure}

The dimensions $a$ and $H$ of the SWG sections containing beam tunnels,
as shown in Fig. \ref{fig:unit_cell_markups}, are selected using
an optimization algorithm which minimizes the integrated frequency
error between the beamline and SWG dispersion curves over the wavenumber
interval of $kd/\pi\in\left[2p,2(p+1)\right]$. This wavenumber interval
corresponds to the frequency range over which amplification is desired
to occur in our paper, as shown in Fig. \ref{SLWG_dispersion_approx}
and Fig. \ref{DualBeamTCSWG_dispersion_approx}. Once suitable $a$
and $H$ dimensions are determined for the SWG with a beam tunnel,
the narrow wall $b$ dimension of our serpentine waveguide way was
chosen to be $b=a/6$ to provide adequate spacing between the SWG
beam tunnel intersections. Of course, in most SWG structures, the
a and b dimensions are rarely close to standard waveguide sizes due
to the need for synchronism with a specific beamline, so waveguide
transitions are needed at the input and output ports to allow connections
for standard waveguide sizes, in addition to radio frequency (RF)
windows to maintain vacuum within the tube. However, for simplicity,
we do not consider such waveguide transitions or windows in our study,
and we only focus on the beam-wave interaction region. The beam tunnel
diameter $d_{t}$ is selected based on the empirical formula from
\cite{nguyen2014design}, 
\begin{equation}
d_{t}=L\alpha\left(1+\left(\frac{L}{2a}\right)^{2}\right)^{-\frac{1}{2}},\label{eq:beam_tunnel_approx-1}
\end{equation}
which minimizes the width of the bandgap caused by the beam tunnel.
The ratio of the beam tunnel radius to the free space wavelength at
the $2\pi$ frequency, $\alpha\simeq0.115$, was used in the design
of our structures as well. The bandgap normally caused by the beam
tunnel is significantly widened due to the additional periodic reactive
loading introduced by coupling slots. Increasing the size of the coupling
slots introduces stronger coupling between the waveguide sections,
but enlarges the bandgap. Therefore, simply having a beam tunnel which
is sufficiently in cutoff appears to be adequate for these kinds of
structures. A square-shaped beam tunnel with side length $d_{t}$
may also be used in place of a conventional cylindrical beam tunnel,
as shown in Fig. \ref{fig:unit_cell_markups}, to make the structures
more compatible with multi-step LIGA (lithographie, galvanoformung,
abformung; German for lithography, electroplating, and molding) processes
\cite{backer1982production}. Traveling wave tube amplifiers with
square beam tunnels may be fabricated using two-step LIGA processes
like in \cite{li2012microfabrication,li2013fabrication,joye2010uv,han2004experimental,park2006feasibility,shin2003novel,li2015uv}.
In the two-step LIGA process, the SWSs are electroformed out of two
symmetric halves which are later bonded together. However, more than
two steps will likely be necessary for our structures due to the coupling
slot lengths differing from the beam tunnel width or differing broad
wall dimensions in each waveguide way. Additionally, the use of a
square beam tunnel may slightly degrade the hot operation of TWTs,
as mentioned in \cite{joye20113d,park2006feasibility,han2004experimental}.
While it is potentially challenging to fabricate such structures using
LIGA fabrication, it should not be significantly more challenging
than it already is for conventional serpentine waveguides fabricated
by two-step LIGA. For example, each additional LIGA step required
for the SLWG and TCSWG structures corresponds to a repetition of procedures
7-13 (lapping/polishing, photoresist attachment, mask, n\textsuperscript{th}
layer alignment, exposure, development, and electroplating) after
procedure 13 shown in \cite{park2006feasibility}.

The dispersion condition of the waveguides (when uncoupled to each
other) utilized by our group to consistently obtain SIPs is to have
two nearly parallel uncoupled (individual-waveguide) modes cross over
a third (individual-waveguide) mode which is nearly perpendicular
to the other two modes on the dispersion diagram, as shown in Fig.
\ref{fig:dispersion_approx}. If a periodic coupling is introduced
between all three of the individual-waveguide modes and the nearly
parallel individual-waveguide modes are in close proximity in the
dispersion diagram, then two phase- and frequency-shifted bandgaps
will form at the intersection points. If the top band-edge of one
bandgap tangentially touches the bottom band-edge of another bandgap,
an inflection point is able to form between these band-edges, as illustrated
in the inset of Fig. \ref{SLWG_dispersion_approx} and Fig. \ref{DualBeamTCSWG_dispersion_approx}.
Varying the proximity of near-parallel individual-waveguide modes
for a given coupling strength directly controls the tilt of the TIP.
Near-parallel individual-waveguide modes which are close to each other
tend to form smooth-TIPs, whereas near-parallel individual-waveguide
modes which are further from each other will typically form alternating-TIPs.
Between smooth-TIP and alternating-TIP conditions, an SIP condition
is expected to exist.

Once the basic dimensions of the SWGs with beam tunnels are established,
coupling slots are positioned between the bends of adjacent waveguide
ways to periodically couple the individual-waveguide modes. The coupling
slots have a vertical thickness $t$ in the y-direction, width $w$
in the z-direction, and length $l$ in the x-direction, as shown in
Fig. \ref{fig:unit_cell_markups}. The length of the coupling slot,
which is also along same axis as the broad wall dimension of the waveguide
ways, is the dimension that strongly controls the evanescent coupling
of modes between waveguide ways. The width of the coupling slot controls
the wave impedance within the slot, and the slot thickness primarily
controls the extent of evanescent decay and phase delay for waves
which are below the slot cutoff frequency. In this paper, we use a
coupling slot length equal to half the $a$ dimension of the SWG section
containing the beam tunnel. The slot width and thickness are arbitrarily
chosen in this paper to demonstrate that SIPs can be attained in our
structures. Larger slot lengths will strengthen mode coupling, but
the dispersion relation of the actual structure will be strongly dissimilar
to the dispersion relation of the individual uncoupled waveguide modes
from before. While large slot lengths and widths enhance mode coupling,
the reflections introduced by the periodic slot reactance in a finite-length
structure may also make the hot structure more susceptible to regenerative
oscillations.

Finally, a small geometric difference is introduced between the waveguide
sections corresponding to near-parallel dispersion curves to control
the frequency and wavenumber spacing between near-parallel dispersion
curves. This directly controls the tilt of the TIP. Large geometric
differences typically result in an alternating-TIP, whereas small
geometric differences make the TIP smooth. For the TCSWG structure,
geometric differences may be introduced as a height difference $\Delta H=\left|H_{t}-H_{b}\right|$
between the top and bottom serpentine sections ($H_{t}$ and$H_{b}$
in Fig. \ref{DualBeamTCSWG_markup}, respectively). For the SLWG structure,
a broad wall dimension difference $\Delta a=\left|a_{t}-a_{b}\right|$
may be introduced between the top and bottom straight waveguide sections
($a_{t}$ and $a_{b}$ in Fig. \ref{SLWG_Markup}, respectively).
Alternatively, geometric differences may be introduced in periodic
capacitive obstacles loading the top and bottom straight waveguides
of the SLWG structure to achieve the same effect, reducing the number
of steps required for LIGA fabrication. However, it may not be desirable
to use periodic obstacles due to the large reflections they introduce
in the top and bottom waveguide sections of the SLWG structure. Conversely,
the use of unloaded parallel waveguides also reduces the complexity
and reflection coefficients of SWS at the cost of more steps with
LIGA fabrication. 

Once all initial structure dimensions are determined, the full unit
cell geometry may be simulated in an eigenmode solver to obtain the
Pierce (interaction) impedance and $\omega-k$ dispersion relation
with real-valued wavenumber, $k$. In this paper, we use the software
CST Studio Suite to obtain the modal dispersion for each periodic
structure. Using simulated dispersion data, the $a$ and $H$ dimensions
of SWG ways containing beam tunnels are then further tuned to recover
beamline synchronism. Adjusting the $a$ dimensions of each serpentine
waveguide shown in Fig. \ref{SLWG_Markup} ($a_{m}$, for the middle
SWG of the SLWG structure; $a_{t}$ and $a_{b}$ for the top and bottom
SWGs of the TCSWG structure) primarily serves to shift each respective
SWG dispersion curve up or down. Adjusting the $H$ dimension primarily
controls the slope of the SWG dispersion curve, which may have changed
due to periodic loading from the slots. When the modal dispersion
curves and TIP are satisfactorily synchronized to the beamline, the
geometric difference between waveguide sections corresponding to parallel
dispersion curves (for example, $\Delta a$ in the SLWG structure
or $\Delta H$ in the TCSWG structure) may then be tuned to adjust
the tilt of the TIP to make it close to an SIP.

\noindent 

\section{Results\label{sec:Results}}

Following the aforementioned design procedures of the previous section,
we obtain the real part of the modal dispersion relation (for the
lossless and cold structures shown, imaginary parts of the dispersion
relation correspond to evanescent modes, e.g. below the cutoff frequency
of the waveguide or in bandgaps where neighboring spatial harmonics
meet on the dispersion diagram) for both SLWG and two-beam TCSWG structures,
as shown in Figs. \ref{SLWG_Dispersion_Tilts} and \ref{DualBeamTCSWG_Dispersion_Tilts},
obtained using the methods shown in Appendix \ref{sec:T-Parameters}
and verified using the eigenmode solver in CST Studio Suite. In the
insets of these figures, we also demonstrate how it is possible to
vary the tilt of the inflection point for three different cases simply
by altering the difference in straight waveguide widths ($a_{t}$
and $a_{b}$ for top and bottom rectangular waveguides, respectively)
for the SLWG structure, or the serpentine waveguide heights ($H_{t}$
and$H_{b}$) of the top and bottom waveguides. The dimensions of each
case are provided in Appendix \ref{sec:Dimensions-used-in}. From
the dispersion relation shown for the SLWG in Fig. \ref{SLWG_Dispersion_Tilts},
the point where the beamline intersects an SIP or smooth-TIP (solid
black and dashed blue curves, respectively) is a backward-wave interaction,
making the SLWG design better suited for use in a BWO rather than
a TWT. While an alternating TIP (magenta dotted curve in the inset
of Fig. \ref{SLWG_Dispersion_Tilts}) might enable forward-wave interactions
in the SLWG structure, the upper and lower band-edges on either side
of the inflection point still pose a significant risk for oscillations.
Backward wave oscillators constructed with the SLWG structure may
also benefit from improved power handling capability compared to a
conventional SWG BWO due to the guided electromagnetic mode being
distributed over a larger cross section in the two lateral waveguides.

From the dispersion relation for the TCSWG structure in Fig. \ref{DualBeamTCSWG_Dispersion_Tilts},
the point where the beamline intersects the inflection point is a
forward wave for the SIP and smooth-TIP (solid black and dashed blue
curves, respectively), making the TCSWG structure a better choice
for TWT designs. While the TCSWG structure shown is designed for velocity-synchronism
with an electron beam at the inflection point, which is inherently
a narrowband phenomenon, this does not mean that the bandwidth of
TWTs built using the TCSWG structure is severely limited. Because
we initially designed these structures with broadband beam-wave synchronization
in mind, there are still broad frequency ranges along dispersion branches
above and below the designed inflection points, where the beamline
is velocity-synchronized to the guided waves. It is still possible
to amplify waves over broad bandwidths as long as the TWT is stable
when the beam is introduced.

Additionally, metal losses and manufacturing errors may perturb the
dispersion relation of an SIP or TIP from its intended design, so
such effects need to be considered in the fabrication and testing
of SLWG and TCSWG structures. Waveguide structures exhibiting EPDs
are known to be quite sensitive to losses and fabrication tolerances\cite{othman2017experimental,abdelshafy2018exceptional,oshmarin2021experimental,nada2020frozen}.
For the example SLWG structure in this paper, the geometric difference
in $\Delta a$ needed to go from an SIP to a smooth/alternating TIP
in the inset of Fig. \ref{SLWG_Dispersion_Tilts} is $30\ \mathrm{\mu m}.$
For the example TCSWG structure, a difference in $\Delta H$ of $50\ \mathrm{\mu m}$
is needed to go from an SIP to a smooth/alternating TIP in the inset
of Fig. \ref{DualBeamTCSWG_Dispersion_Tilts}. The work of Li, \emph{et.
al.} \cite{li2013fabrication} shows that it is possible to fabricate
serpentine waveguide structures with $6\ \mathrm{\mu m}$ depth tolerance
(for the broad-wall dimension $a$ of our structures) and $2\ \mathrm{\mu m}$
width tolerance between sidewalls using ultraviolet LIGA (UV-LIGA)
fabrication techniques. A fabrication tolerance of $6\ \mathrm{\mu m}$
(20\% of the difference in $\Delta a$ needed to go from an SIP to
a smooth/alternating TIP shown in Fig. \ref{SLWG_Dispersion_Tilts}
for the SLWG structure) is acceptable to obtain a SLWG structure that
exhibits a nearly-stationary inflection point. For the TCSWG structure,
a tolerance of $2\ \mathrm{\mu m}$ for sidewall widths (4\% of the
difference in $\Delta H$ needed to go from an SIP to a smooth/alternating
TIP shown in Fig. \ref{DualBeamTCSWG_Dispersion_Tilts} for the TCSWG
structure) is even better. Though, of course, these tolerances may
be relaxed for designs that have an SIP/TIP at lower frequencies.

Due to the neighboring upper and lower band edges in proximity to
the inflection point, one must also carefully ``aim'' the beamline,
which is directly controlled by the accelerating voltage of the electron
gun, to avoid striking dispersion branches that have zero group velocity,
such as the band edge, as it can lead to instability \cite{hung2015absolute,zhang2016absolute,antoulinakis2018absolute,ang1998absolute}.
Because there are upper and lower band edges at frequencies close
to the inflection point, the tuneability of the beam voltage is limited.
For instance, in Fig. \ref{SLWG_Dispersion_Tilts}, neglecting space
charge (i.e. at low beam currents), the average beam velocity can
only be varied by approximately $u_{0}=0.200c\pm0.001c$ to avoid
striking the neighboring upper or lower band edges on other dispersion
branches. This beam velocity range corresponds to an approximate kinetic
equivalent voltage tuneable range of $V_{0}=10.54\pm0.11\ \mathrm{kV}$
from the relativistic relation $V_{0}=c^{2}/\eta\left[1-\left(u_{0}/c\right)^{2}\right]^{-1/2}$in
\cite{gilmour1994principles_ch3}, where $\eta$ is the charge-to-mass
ratio of an electron at rest. For the case of the TCSWG structure
we show, the tuneable range of the beam velocity equivalent kinetic
voltage is better due to the neighbouring upper/lower band edges near
the inflection point being separated at higher/lower frequencies,
respectively, as can be observed in Fig. \ref{DualBeamTCSWG_Dispersion_Tilts}.
However, there is still a risk of oscillations at the lower band edge
corresponding to a frequency of approx. 133 GHz for the structure
shown, so the tuneable range of beam velocity for the TCSWG structure
is $u_{0}=0.300c\pm0.002c$, which corresponds to an approximate beam
voltage range of $V_{0}=24.67\pm0.36\ \mathrm{kV}$.

One must also consider how the electron beam perturbs the inflection
point in the hot dispersion relation for the three-mode synchronization
regime, as was explained in \cite{yazdi2017new}. When an electron
beam is introduced to the system, the SIP or TIP which existed in
the cold dispersion relation will be deformed in the hot dispersion
relation. The amount that the electron beam perturbs the inflection
point depends primarily on the dc beam current density, with higher
currents causing larger perturbations to the inflection point in the
hot dispersion relation.

\begin{figure}
\subfloat[]{\includegraphics[width=0.9\columnwidth]{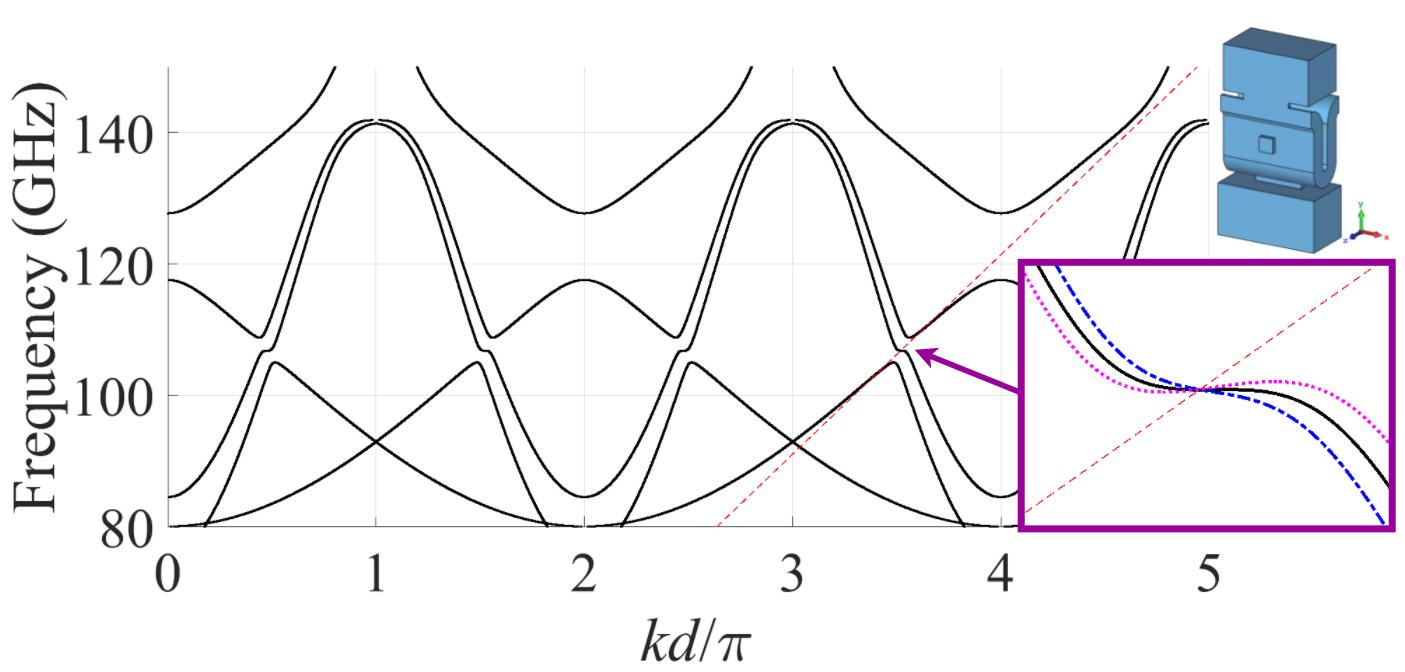}\label{SLWG_Dispersion_Tilts}}\medskip{}
\subfloat[]{\includegraphics[width=0.9\columnwidth]{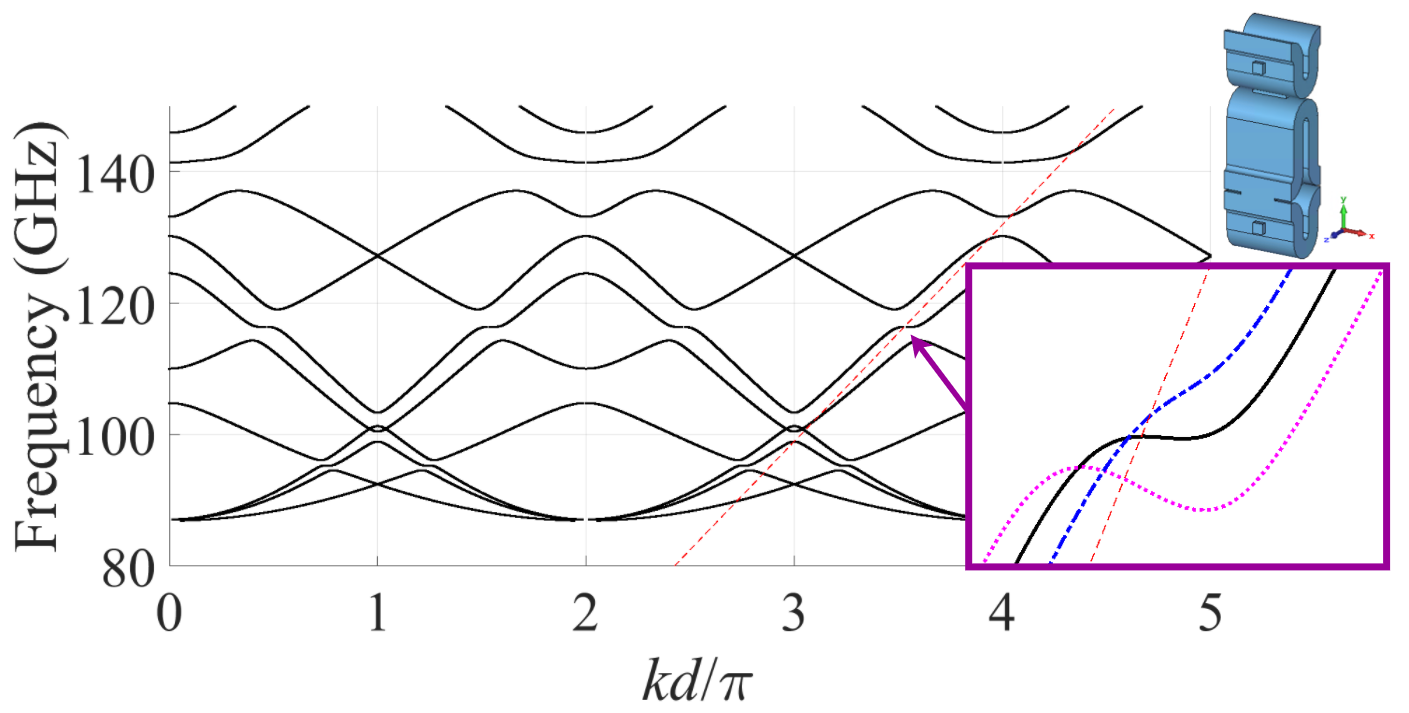}

\label{DualBeamTCSWG_Dispersion_Tilts}}

\caption{Modal dispersion diagrams for (a) SLWG unit cell, and (b) TCSWG unit
cell, with beamline (red dashed). The insets show a smooth-TIP, alternating-TIP,
and SIP in blue (dashed line), magenta (dotted line), and black (solid
line), respectively for each structure. Only wavenumbers which are
purely real are shown. The dispersion diagrams in the figure are obtained
using the methods shown in Appendix \ref{sec:T-Parameters}, and the
dispersion relations were verified using the full-wave eigenmode solver
of CST Studio Suite. Tilting of the inflection point is achieved by
fine-tuning one or more of the structure dimensions. Dimensions for
the smooth-TIP, alternating-TIP, and SIP are available in Appendix
\ref{sec:Dimensions-used-in}.}
\end{figure}

\begin{figure}
\subfloat[]{\includegraphics[width=1\columnwidth]{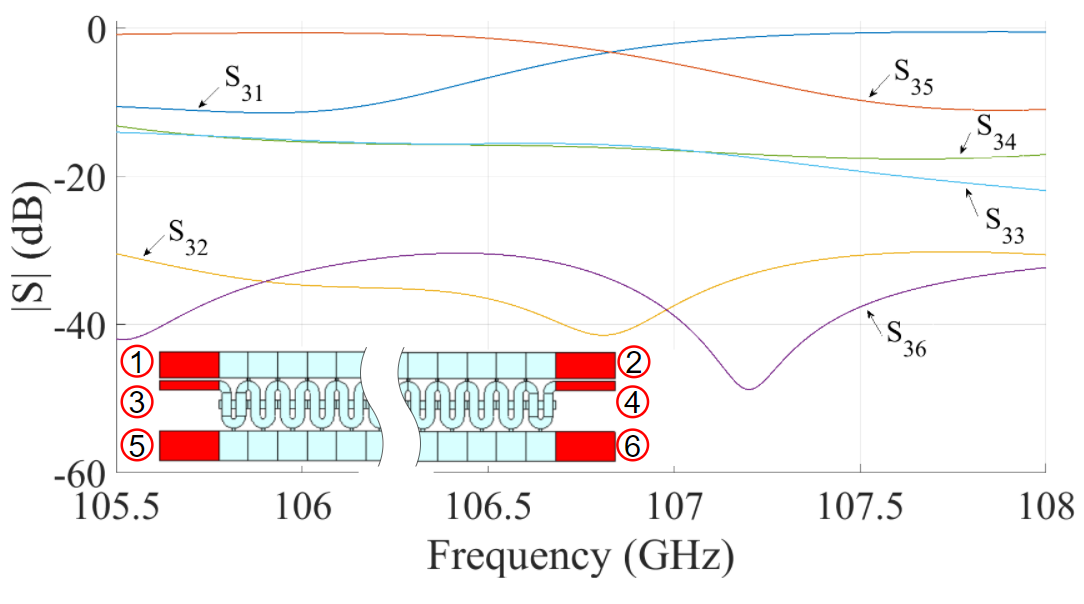}

\label{SLWG_SParams_PortNumbering}}\medskip{}
\subfloat[]{\includegraphics[width=1\columnwidth]{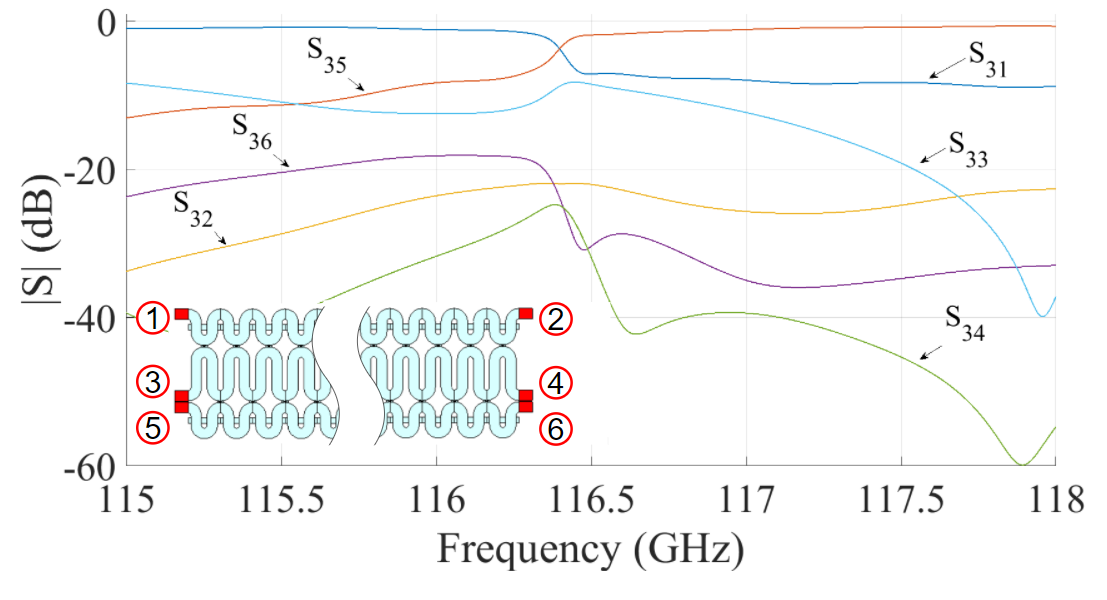}

\label{DualBeamTCSWG_SParams_PortNumbering}}

\caption{SWS with 32 unit cells and S parameters for (a) SLWG and (b) TCSWG
in the vicinity of the SIP with port numbering scheme. Port de-embedding
is shown in red. Scattering parameters are calculated using the method
explained in Appendix B.}
\end{figure}

Similar to other three-way waveguide power dividers and combiners
explored in papers such as \cite{fonseca2013design,gardner1997mode,kumar2017compact},
our structures also exhibit directional coupler-like behavior at its
SIP/TIP frequencies, as demonstrated with a finite-length structure
of 32 unit cells, shown in Figs. \ref{SLWG_SParams_PortNumbering}
and \ref{DualBeamTCSWG_SParams_PortNumbering}. The S parameters of
the finite length structure were calculated using the methods explained
in Appendix \ref{sec:T-Parameters}. The port numbering scheme for
our structures is that the input ports at the electron-gun end of
the structure (on the left) are odd-numbered from top to bottom, and
output ports at the collector end of our structure (on the right)
are even-numbered from top to bottom. While it is highly important
to consider the effect of waveguide transitions and RF windows, our
study focuses primarily on the interaction region of linear beam tubes,
so for brevity we do not consider the effect of input/output coupling
structures; i.e. we only consider the S-parameters at reference planes
between the SWS and where an RF window would be placed in a fabricated
device. This directional coupler-like behavior enables distributed
power extraction (DPE) which can be directed either backward toward
the cathode-end of the structure or forward toward the collector-end
of the structure. However, only forward-directive DPE may be desired
for amplification, due to the potential risk of regenerative oscillations
introduced by amplified waves returning to the electron gun-end of
the structure, like in \cite{bhattacharjee2004folded}. The introduction
of distributed power extraction for linear beam tubes was necessary
to conceive the degenerate (or exceptional) synchronization in the
hot systems studied in \cite{mealy2019exceptional,mealy2021high2,mealy2021high}.
For the TCSWG structure, increasing the path length of the middle
SWG can allow forward-directive DPE to occur at certain frequencies
and for power to be extracted in the top, middle, and bottom SWG outputs.
However, there may still be SIP/TIP frequencies where backward-directive
DPE continues to occur. Shifting the frequency above or below the
SIP/TIP in the vicinity of the inflection point directly controls
whether the top or bottom SWG section contributes more power to the
output of the middle SWG, as can be seen from the scattering parameters.
This dual beam TCSWG structure may also be excited either from the
middle SWG input or top and bottom SWG inputs to achieve amplification
and DPE at the SIP/TIP frequency if coupling is sufficient. Increasing
the size of coupling slots enhances DPE, however this also exacerbates
reflections in the finite-length structure and increases risk of BWO.
Because these structures can still be designed to be well-matched
with small coupling slots, longer finite-length structures and higher
beam currents may potentially be used before unstable BWO occurs in
PIC simulations and hot testing.

\begin{figure}
\includegraphics[width=0.9\columnwidth]{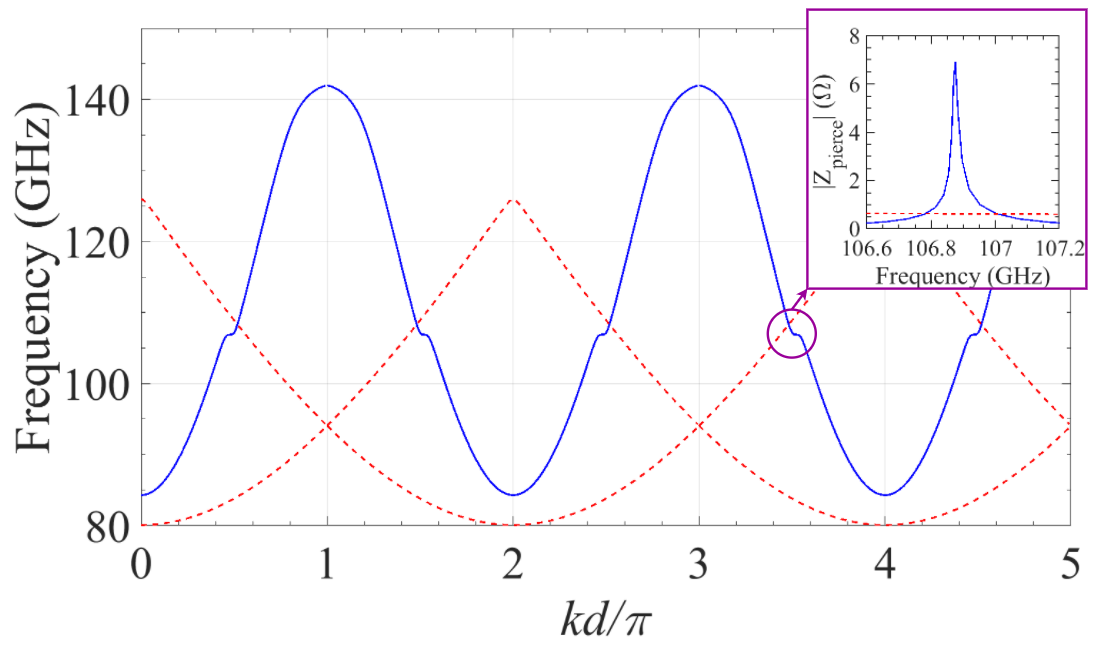}

\caption{Pierce (interaction) impedance (solid blue curve in inset) for the
SLWG calculated from the field along the beam tunnel for the eigenmode
with a TIP, associated to the circled region in the dispersion diagram
(solid blue curve in main figure). Nearly-stationary TIP with negative
fundamental $k$ (i.e., in the fundamental Brillouin zone defined
here as $kd/\pi\in[-1,1]$) is circled for the $p=2$ spatial harmonic.
The Pierce impedance at the TIP is compared to the Pierce impedance
associated with the eigenmode of an individual serpentine waveguide
with coupling slots and straight waveguide geometries removed (red
dashed). Specifically, the Pierce impedance was evaluated over the
phase interval $kd/\pi\in[-0.53,-0.43]$ for the TIP at the appropriate
spatial harmonic $p=2$.The dispersion diagram of the individual SWG
mode (dashed red) is also shown in the main figure. The dispersion
curves in the main figure were found using the eigenmode solver of
CST Studio Suite.}
\label{SLWG_Pierce_impedance}
\end{figure}

Finally, we compute the Pierce impedance as discussed in Section \ref{sec:Cold-Stationary-Inflection}
for a fourth case of the SLWG structure in the vicinity of a nearly-stationary
smooth-TIP, which has dispersion relation similar to the black curve,
but the inflection point is not as not as tilted as the blue curve,
shown in Fig. \ref{SLWG_Dispersion_Tilts}, with dimensions provided
in Appendix \ref{sec:Dimensions-used-in}. We demonstrate the benefit
of using nearly-stationary TIPs to enhance the Pierce impedance of
serpentine-like structures, as shown in Fig. \ref{SLWG_Pierce_impedance}.
We also compare the Pierce impedance of the SLWG (solid blue line)
to the the Pierce impedance of a conventional individual SWG (red
dotted line) (i.e., the serpentine of the SLWG structure, with removed
coupling slots and straight waveguide ways). We find that the pierce
impedance of the full SLWG structure is several times higher than
a conventional simple serpentine waveguide at the frequency corresponding
to a nearly-stationary inflection point. We also note that, while
the interaction impedance of the SLWG appears quite small relative
to the simple SWG at frequencies beyond the inflection point, the
interaction impedance is comparable that of an SWG on other higher/lower
frequency branches of the SLWG's dispersion diagram, which are not
shown. 

\noindent 

\begin{figure}
\includegraphics[width=0.9\columnwidth]{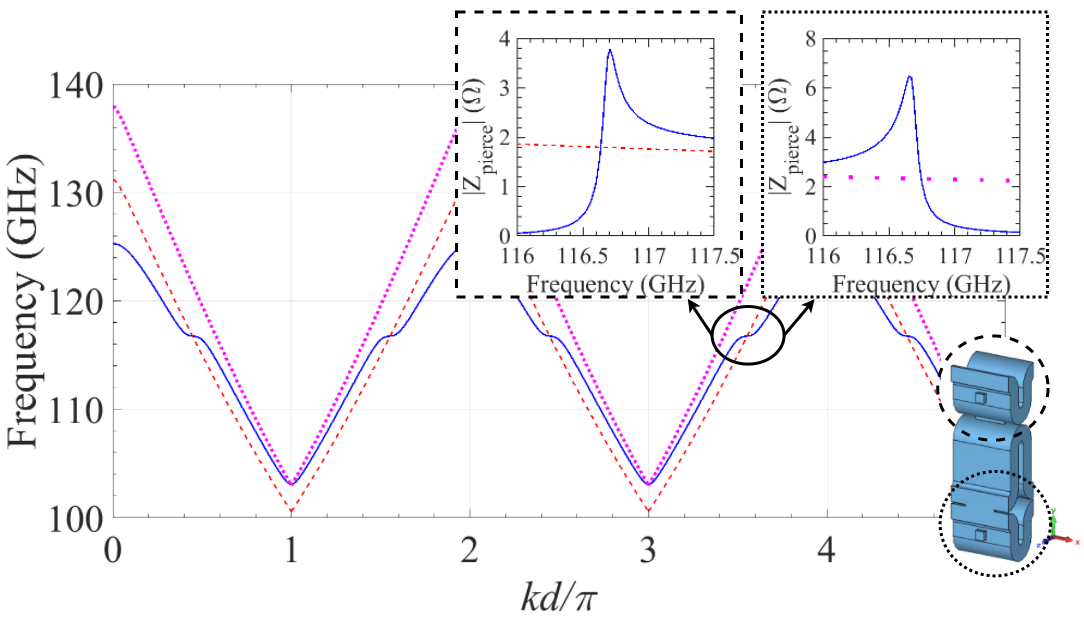}

\caption{Pierce (interaction) impedance (solid blue curves) for the top beam
tunnel (inset with black dashed border) and bottom beam tunnel (inset
with black dotted border) of the TCSWG structure for the eigenmode
with a TIP, associated to the circled region in the dispersion diagram
showing a nearly-stationary TIP with negative fundamental $k$ in
the $p=2$ spatial harmonic (solid blue curve in main figure). The
Pierce impedance at the TIP is compared to the Pierce impedance associated
with the eigenmodes of the individual serpentine waveguides at the
top of the structure (dashed red) and at the bottom of the structure
(dotted magenta) when the coupling slots and other waveguide sections
are removed. Specifically, the Pierce impedance in both tunnels of
the TCSWG structure were calculated over the phase interval $kd/\pi\in[-0.56,-0.35]$
for the TIP at the appropriate spatial harmonic $p=2$. The dispersion
branches of the individual SWG modes associated to the top SWG (dashed
red) and bottom SWG (dotted magenta) are also shown in the main figure.
The dispersion curves of the main figure were found using the eigenmode
solver of CST Studio Suite.}
\label{TCSWG_Pierce_impedance}
\end{figure}

Similarly, we compute the Pierce impedance for the a fourth case of
the dual-beam TCSWG structure in the vicinity of the nearly-stationary
smooth-TIP, which has a dispersion similiar to the black curve, but
not as tilted as the blue curve shown in the inset of Fig. \ref{DualBeamTCSWG_Dispersion_Tilts},
with dimensions provided in Appendix \ref{sec:Dimensions-used-in}.
We demonstrate that the nearly-stationary TIP can be used to ehance
the Pierce impedance in both beam tunnels, as shown in Fig. \ref{TCSWG_Pierce_impedance}.
We compare the Pierce impedance of the TCSWG structure (solid blue
line) to the Pierce impedance of a conventional individual SWG (i.e.
with coupling slots and adjacent waveguide sections removed) for each
respective beam tunnel (dashed red for the top SWG, dotted magenta
for the bottom SWG). We find that, just like with the SLWG structure,
the Pierce impedance is several times higher than a conventional SWG
at the frequency corresponding to the nearly-stationary inflection
point. Interestingly, below the frequency of the TIP, the interaction
impedance in the lower beam tunnel is higher than in the upper beam
tunnel, whereas the opposite occurs at frequencies above the TIP.
Since glide symmetry is slightly broken due to the top and bottom
SWGs having different $H_{t}$ and $H_{b}$ dimensions, respectively,
the dispersion branches of the individual SWGs (red dashed line for
the top SWG and magenta dotted line for the bottom SWG) are dissimilar.
Due to broken glide symmetry, the peak values of interaction impedance
in the top and bottom tunnels are also different at the inflection
point. If an electron beam is introduced to the SLWG or TCSWG structures
and the beam is velocity synchronized to the SIP/TIP, we say that
the electron beam is synchronized to three degenerate modes, i.e.,
we have three-mode synchronization, as was described in \cite{yazdi2017new}.
Under the three-mode synchronization regime, the Pierce gain parameter
$C$ will also become larger than that of a conventional SWG TWT due
to the enhanced interaction impedance. 

\noindent 

\section{Conclusion}

We have showcased two novel dispersion-engineered three-way SWSs for
use in linear electron beam devices: the SLWG and TCSWG geometries.
Such geometries are capable of exhibiting SIPs or TIPs in their dispersion
relations, and larger Pierce (interaction) impedance than that of
a conventional serpentine waveguide at the frequency corresponding
to the inflection point. Using our design methodology, we were able
to demonstrate simple conditions which enable one or more SIP/TIP
to occur in a three-way waveguide periodic structure once weak periodic
coupling is introduced between individual waveguides. We have shown
the first known example of a millimeter-wave SWS for linear beam tubes
which exhibits stationary or nearly-stationary inflection points in
its dispersion relation. A previous example of a waveguide which exhibits
an SIP at radio frequencies was demonstrated using microstrip technology
in \cite{nada2020frozen}, and was the inspiration for this paper.

What is of interest in both of our introduced structures is that the
group velocity in the vicinity of the SIP/TIP may be easily controlled
by slightly breaking glide-symmetry in our geometries. With weak coupling
between waveguides, the dispersion relation of the introduced structures
will not be significantly different from the dispersion relations
of the individual (uncoupled) waveguides. Due to the ``three-mode
synchronization'' regime, the Pierce impedance, and consequently
the Pierce gain parameter, can be drastically enhanced over narrow
bandwidths near the SIP/TIP, when compared to a conventional serpentine
waveguide commonly used for millimeter-wave TWTs and BWOs. We believe
that the introduction of the three-mode synchronization regime in
such structures may enable the design of more efficient, compact linear
beam tubes, as it was speculated in \cite{yazdi2017new}. There is
a lot of room for improvement if one wishes to focus on improving
the interaction impedance or bandwidth further using SIPs. However,
we believe that the design methodology shown in this paper is still
useful for designing realistic millimeter-wave SWSs with inflection
points in their dispersion relation for use in linear beam tubes.Additionally,
we have shown how to obtain TIPs with either backward or forward mode
interactions by simply by varying how much glide-symmetry is broken
in our structures, potentially enabling the design of either BWOs
or TWTs from the same initial geometry. We have also showcased the
directional coupler-like properties of both the TCSWG and SLWG structures
at frequencies near-to the SIP/TIP. This property may be exploited
in the design of specialized TWTs or BWOs which can be simultaneously
excited from multiple ports or simultaneously drive multiple loads.

\appendices{}

\section{Dimensions used in Figures\label{sec:Dimensions-used-in}}

\begin{table}
\caption{SLWG and TCSWG dimensions used in Figures \ref{SLWG_Dispersion_Tilts}
and \ref{DualBeamTCSWG_Dispersion_Tilts}. Dimensions are in $\mathrm{mm}$
unless specified otherwise\label{tab:SLWG-and-TCSWG}}

\centering{}%
\begin{tabular}{|>{\centering}p{0.25in}|>{\centering}p{0.75in}|>{\centering}p{0.75in}|>{\centering}p{1in}|}
\hline 
 & Dimensions used for SWLG & Dimensions used for TCSWG & Notes\tabularnewline
\hline 
$a_{t}$, $b_{t}$ & $1.879-\Delta a/2$, $a_{t}/2$ & 1.726, 0.432 & Cross-section dimensions for top waveguide\tabularnewline
\hline 
$a_{m}$, $b_{m}$ & 1.879,$a_{m}/6$ & 1.726, 0.432 & Cross-section dimensions for middle waveguide\tabularnewline
\hline 
$a_{b}$, $b_{b}$ & $1.879+\Delta a/2$, $a_{b}/2$ & 1.726, 0.432 & Cross-section dimensions for bottom waveguide\tabularnewline
\hline 
\emph{l} & 0.940 & 0.863 & Rectangular slot length along x-dimension\tabularnewline
\hline 
\emph{W} & 0.100 & 0.259 & Rectangular slot width along z-dimension\tabularnewline
\hline 
\emph{t} & 0.100 & 0.050 & Rectangular slot thickness between waveguide bends\tabularnewline
\hline 
$d_{t}$ & 0.282 & 0.259 & Width of square beam tunnel\tabularnewline
\hline 
\emph{d} & 0.989 & 1.363 & Unit cell pitch\tabularnewline
\hline 
$H_{t}$, $H_{m}$, $H_{b}$ & $H=$0.784 & $0.472+\Delta H/2$, 1.312, $0.472-\Delta H/2$ & Height of straight sections for top, middle, and bottom SWGs. SLWG
only has one \emph{H} dimension\tabularnewline
\hline 
$u_{0}$/c & 0.200 & 0.300 & Average electron beam velocity normalized to the speed of light\tabularnewline
\hline 
\end{tabular}
\end{table}

For the dispersion of the individual-waveguide ways of the SLWG structure
in Fig. \ref{SLWG_dispersion_approx} (i.e, without coupling), the
dimensions are identical to those in Table \ref{tab:SLWG-and-TCSWG}
with $\Delta a=0.1$\ mm. , with the exception of $H=0.843$\ mm,
which differs from the final $H$ dimension used for shown in Fig.
\ref{SLWG_Dispersion_Tilts}. Note that, unlike the TCSWG structure,
the SLWG structure only has one $H$ dimension.

For the dispersion of the individual waveguide ways of the TCSWG structure
in Fig. \ref{DualBeamTCSWG_dispersion_approx}, the dimensions are
identical to those in Table \ref{tab:SLWG-and-TCSWG} with the exception
of the normalized beam velocity being $u_{0}/c=0.20$ and the top,
middle, and bottom SWG heights being $H_{t}=0.518$\ mm, $H_{m}=1.161$\ mm,
and $H_{b}=0.418$\ mm, respectively.

Table I reports the dimensions for Fig. \ref{SLWG_Dispersion_Tilts}
and Fig. \ref{DualBeamTCSWG_Dispersion_Tilts} as follows. In the
inset of Fig. \ref{SLWG_Dispersion_Tilts}, we have the dimensions
used in the SLWG column of Table \ref{tab:SLWG-and-TCSWG} , with
$\Delta a=0.21$\ mm for the SIP (black solid lines), $\Delta a=0.24$\ mm
for the alternating-TIP (group velocity alternates in sign at frequencies
slightly above or below TIP frequency) (magenta dotted line), and
$\Delta a=0.18$\ mm for the smooth-TIP (group velocity does not
change sign at frequencies slightly above or below TIP frequency)
(blue dashed line).

In the inset of Fig. \ref{DualBeamTCSWG_Dispersion_Tilts}, we have
the dimensions used in the TCSWG column of Table \ref{tab:SLWG-and-TCSWG},
with $\Delta H=0.15$\ mm for the SIP (black solid line), $\Delta H=0.10$\ mm
for the smooth-TIP (blue dashed line), and $\Delta H=0.20$\ mm for
the alternating-TIP (magenta dotted line).

For the Pierce impedance plot of Fig. \ref{SLWG_Pierce_impedance},
we use the dimensions of the SWLG column of Table \ref{tab:SLWG-and-TCSWG},
with $\Delta a=0.20$\ mm, which provides an smooth-TIP that is almost
stationary (i.e., it is very close to an SIP). For the Pierce impedance
plot of Fig. \ref{TCSWG_Pierce_impedance}, we use the dimensions
of the TCSWG column of Table \ref{tab:SLWG-and-TCSWG}, with $\Delta H=0.125$\ mm,
which provides a smooth-TIP that is almost stationary. We simulated
the SLWG and TCSWG structures in the eigenmode solver of CST Studio
Suite 2019 with a step size of $0.5^{\circ}$ in the boundary phase
and a tetrahedral mesh setting of 50 cells per max box edge. We found
that the peak value of the interaction impedance at an SIP or nearly-stationary
TIP for both structures depends greatly on mesh density and step size
of boundary phase.

\section{T Parameters\label{sec:T-Parameters}}

Using frequency-domain simulations, it is possible to rapidly and
accurately compute the complex dispersion and approximate finite-length
S-parameters of a periodic structure without an eigenmode solver using
the frequency-dependent scattering parameters (S parameters) of a
single unit cell having separable pairs of ports for each way of the
multi-way waveguide structure. The benefit of computing dispersion
in this manner is that it is possible to obtain both real and imaginary
solutions for the Bloch wavenumber, $k$, whereas it is only possible
to obtain the real part of the Bloch wavenumber using the eigenmode
solver of CST Studio Suite. For lossless and cold structures, the
imaginary part of the Bloch wavenumber corresponds to evanescent modes,
e.g. below the cutoff frequency of the waveguide or in bandgaps that
form where neighboring spatial harmonics meet. It is also notably
faster to obtain the unit cell S parameters using the frequency domain
solver than it is to directly obtain the modal dispersion from a phase
sweep of the periodic boundary in the eigenmode solver. Using this
method, it was possible for us to tune the geometry of our structures
to obtain beamline synchronism and desired inflection point tilt in
a reasonable time frame. The real dispersion obtained by this method
was found to be in excellent agreement with CST Studio Suite's eigenmode
solutions.

The method we use to obtain the complex dispersion and approximate
finite-length S parameters of our periodic structures involves converting
between S parameters and scattering transmission matrices (T parameters)
in intermediate steps. T parameters may be directly obtained through
algebraic manipulation of each S parameter, as both are defined in
terms of the same $a$- and $b$- waves.

\begin{equation}
\left[\begin{array}{c}
b_{1}\\
b_{2}\\
b_{3}\\
b_{4}\\
b_{5}\\
b_{6}
\end{array}\right]=\underline{\mathrm{\mathbf{S}}}\left[\begin{array}{c}
a_{1}\\
a_{2}\\
a_{3}\\
a_{4}\\
a_{5}\\
a_{6}
\end{array}\right]\iff\left[\begin{array}{c}
b_{2}\\
a_{2}\\
b_{4}\\
a_{4}\\
b_{6}\\
a_{6}
\end{array}\right]=\underline{\mathrm{\mathbf{T}}}\left[\begin{array}{c}
a_{1}\\
b_{1}\\
a_{3}\\
b_{3}\\
a_{5}\\
b_{5}
\end{array}\right]\label{eq:MatrixFormulation}
\end{equation}
By converting our frequency-dependent S parameters to T parameters
at each frequency for the unit cell, it is possible to solve the following
Floquet-Bloch eigenvalue problem at each frequency to obtain the complex
dispersion diagram for the modes of the periodic structure \cite{abdelshafy2018electron,othman2017theory,tamma2014extension},

\[
\underline{\mathrm{\mathbf{T}}}(\mathit{z+d,z})\boldsymbol{\Psi}(\mathit{z})=e^{-jkd}\boldsymbol{\Psi}(\mathit{z}),
\]
where, $\boldsymbol{\Psi}$ is the complex state vector for the unit
cell composed of $a$ and $b$ waves at each port, as shown in Eqn.
\ref{eq:MatrixFormulation} and Fig. \ref{SLWG_SParams_PortNumbering}
and Fig. \ref{DualBeamTCSWG_SParams_PortNumbering}, and $d$ is the
unit cell pitch. In other words, the wavenumbers of the fundamental
spatial harmonic, may be evaluated directly from the eigenvalues $\lambda=\exp(-jkd)$
of the T matrix through the relation

\[
k=\frac{-\ln\left(\lambda\right)}{jd}.
\]
The modal dispersion diagrams in Fig. \ref{SLWG_Dispersion_Tilts}
and Fig. \ref{DualBeamTCSWG_Dispersion_Tilts} (only showing the branches
with purely real $k$) are calculated from the eigenvalue problem
shown above and are verified using the eigenmode solver of CST Studio
Suite. It is also possible to estimate the S parameters of our periodic
structure with finite length by cascading T matrices and converting
the resultant parameters back into S parameters using the same algebraic
manipulation as before \cite{egan2004practical_ch2}. This is how
the finite-length S parameters shown in Fig. \ref{SLWG_SParams_PortNumbering}
and Fig. \ref{DualBeamTCSWG_SParams_PortNumbering} were computed.
This method may be readily generalized for 2\textit{N}-port periodic
structures. It is also possible to solve the eigenvalue for spatial
harmonics other than the fundamental, as the solutions $k_{n}$ are
periodic as $k_{n}=k+2\pi n/d$, with $n=0,\pm1,\pm2,...$.

\begin{figure}
\subfloat[]{\includegraphics[width=1.75in]{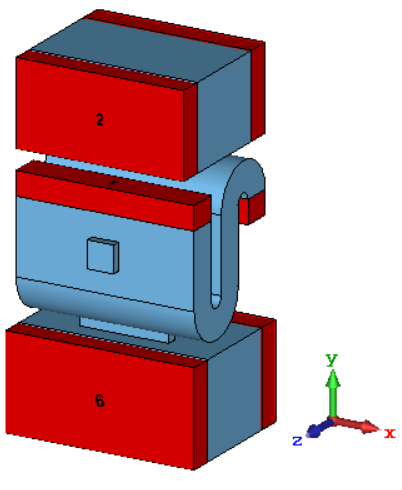}

\label{SLWG_T_param_model}}\subfloat[]{\includegraphics[width=1.75in]{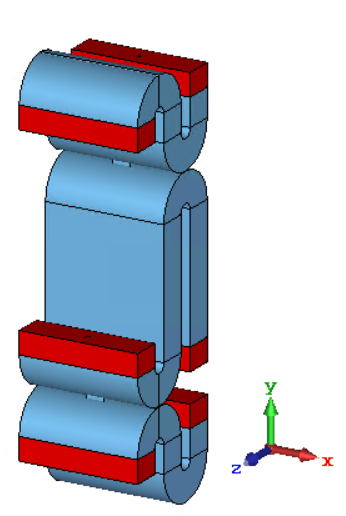}

\label{DualBeamTCSWG_T_param_model}}

\caption{(a) SLWG and (b) dual-beam TCSWG vacuum unit cells used for calculating
the T parameters from the single-cell S parameters. Port reference
planes are shifted $d/4$ along each waveguide path to prevent ports
from being placed too close to the coupling slots. Port de-embedding
is shown in red. Knowledge of the T matrix allows for fast and accurate
calculation of the scattering parameters of multi-way waveguides with
several unit cells.}
\end{figure}

Due to the presence of periodic coupling slots in both of our unit
cell designs, our single-cell frequency domain model for our structures
must be slightly modified in order to avoid placing ports at coupling
slots or in waveguide bends. Our unit cells were modified for the
frequency domain solver by horizontally shifting the reference planes
of all ports by $d/4$ along each waveguide path and adding de-embedding
to the ports to account for small reflections caused by transitions
between straight waveguide sections and waveguide bends and slots,
as shown in Fig. \ref{SLWG_T_param_model} and Fig. \ref{DualBeamTCSWG_T_param_model}.
This modification does not appear to significantly affect the dispersion
relation of the periodic structure under study.

\section*{Acknowledgments}

This material is based upon work supported by the GAANN Fellowship,
by the Air Force Office of Scientific Research under Award FA9550-18-1-0355
and under the MURI Award FA9550- 20-1-0409 administered through the
University of New Mexico. We thank Dassault Systèmes for providing
CST Studio Suite, which has been instrumental in this work. The authors
would also like to thank our colleague, Mr. Miguel Saavedra, for his
assistance in fixing our interaction impedance calculation.


\end{document}